\documentstyle[aps,preprint,prb,epsf]{revtex}
\begin{document}
\newcommand{\zmt}{Zn$_{1-x}$Mn$_x$Te}
\newcommand{\zms}{Zn$_{1-x}$Mn$_x$Se}
\newcommand{\zfs}{Zn$_{1-x}$Fe$_x$Se}
\newcommand{\zcs}{Zn$_{1-x}$Co$_x$Se}
\draft\title{X-ray    absorption     spectroscopy    study     of    diluted     magnetic 
semiconductors:\\Zn$_{1-x}$M$_x$Se (M = Mn, Fe, Co) and Zn$_{1-x}$Mn$_x$Y (Y = 
Se, Te)}
\author {Kwanghyun Cho, Hoon Koh, and S.-J. Oh\cite{address1}}
\address{Department of Physics \&  Center for Strongly  Correlated Materials Research, 
Seoul National University, Seoul 151-742, Korea}
\author {Hyeong-Do Kim and Moonsup Han}
\address{Department of Physics, University of Seoul, Seoul130-743, Korea}
\author {J.-H. Park}
\address{Department of Physics, Pohang University of  Science and Technology, Pohang 
790-784, Korea}
\author {C. T. Chen}
\address{Synchrotron Radiation Research Center, Hsinchu Science-based  Industrial Park, 
Hsinchu 300, Taiwan, Republic of China}
\author {Y. D. Kim}
\address{Department of Physics, Kyung Hee University, Seoul 130-701, Korea}
\author {J.-S. Kim}
\address{Department of   Physics, Sook-Myung   Women's University,  Seoul  140-742, 
Korea}
\author {B. T. Jonker}
\address{Naval Research Laboratory, Washington, DC 20375-5343}
\date{Received \hspace*{30mm}}
\maketitle
\begin{abstract}
We have investigated 3$d$ electronic states  of doped transition metals in  II-VI diluted 
magnetic   semiconductors,    Zn$_{1-x}$M$_x$Se   (M   =   Mn,    Fe,   Co)   and 
Zn$_{1-x}$Mn$_x$Y (Y =  Se, Te),  using the transition-metal  $L_{2,3}$-edge X-ray 
absorption spectroscopy (XAS) measurements. In order to  explain the XAS spectra, we 
employed a tetragonal cluster  model calculation, which includes  not only the full  ionic 
multiplet structure but also  configuration interaction (CI). The  results show that CI  is 
essential  to  describe  the   experimental spectra   adequately,  indicating  the  strong 
hybridization between the  transition metal  3$d$ and  the ligand  $p$ orbitals.  In the 
study of Zn$_{1-x}$Mn$_x$Y (Y = Se, Te), we also found considerable spectral change 
in the Mn $L_{2,3}$-edge XAS spectra for  different ligands, confirming the importance 
of the hybridization effects in these materials.
\end{abstract}

\section{Introduction}
There has been active studies on diluted magnetic semiconductors  (DMS's) for the last 
two decades because of  their unique magneto-optical and  magneto-transport properties 
such as giant negative  magnetoresistance, extremely large  electronic $g$ factor,  large 
Faraday rotation,   etc.\cite{Furdyna,Review} Recently  the  interest  has  surged again 
because of  their possible  applications  to ``spintronics''  based on   the semiconductor 
technology.\cite{Nature} DMS's, which are made by substitution of small amount of late 
3$d$ magnetic transition-metal (TM)  atoms such as Mn,  Fe, and Co ions  into cation 
sites in  the II-VI  or  III-V compound  semiconductors, involve    both semiconductor 
physics and magnetism.  The band  gap energy turns  out to  change with the  doping 
concentration.\cite{YDKim} Moreover,   in some  DMS's, the   variation of  the doping 
concentration was found to induce magnetic  phase transitions, such as paramagnetic  to 
spin glass  phase and  spin glass  to antiferromagnetic   phase,\cite{Furdyna} and even 
ferromagnetism   can    also   be   induced    by   injecting   carriers    or   photon 
irradiation.\cite{Ohno} 
  
Here our studies are mainly focused  on the electronic structure of II-VI  based DMS's. 
The II-VI compounds such as ZnSe  and ZnTe have a zinc-blende crystal  structure, in 
which the cation Zn$^{2+}$  ions are under the  tetrahedral $T_d$ site symmetry.  The 
TM doping induces  merely small  changes of  the lattice-constants within  the crystal 
structure. The $(4sp)^2$ electrons  of the TM atoms,  which act like the  Zn $(4sp)^2$ 
electrons, participate  in the  ligand $p$-Zn  $4sp$  bonding-antibonding states.  These 
states, which correspond to the respective valence band and conduction band of the host 
semiconductor, are hardly  affected by  the doping\cite{Furdyna,Jonker}. Meanwhile,  the 
TM 3$d$ electrons provide rather localized additional band  states, which is expected to 
be located energetically in the wide band gap of the bonding-antibonding states.  Optical 
absorption studies of DMS's showed the intra-atomic $d$-$d$ transitions which can be 
understood by   the 3$d$  Coulomb  multiplets excited   from the  corresponding  ionic 
high-spin ground state.\cite{Mizokawa,dd} In  spite of the ionic  characteristics, the TM 
3$d$ electrons are known to  make strong covalent bonding  with ligand $p$ electrons. 
Furthermore,  this  covalent  bonding,   which is   often represented   by  ``$p$-$d$'' 
hybridization, is expected to play an important role in  the interesting magneto-transport 
and   magneto-optical  properties   such  as   giant  Faraday   rotation  and   Zeeman 
splitting.\cite{Furdyna} 

In order to understand  the electronic structure  of these DMS's,  several photoemission 
spectroscopy measurements have been  performed, mostly utilizing  $3p \rightarrow 3d$ 
resonance phenomena near  the TM 3$p$  absorption edge.\cite{PES,CI} This  technique 
enables us to separate out the TM 3$d$ contribution to the electronic structure, and one 
can extract a  sort of  the 3$d$  partial spectral  weight. The  resulting 3$d$  spectral 
weight was found to be distributed in a very wide range of the valence band, no matter 
what the doping  concentration is, indicating  the strong  $p-d$ hybridization of  ligand 
$p$ band and the TM 3$d$ orbitals. Meanwhile,  the 3$d$ spectral line shape does not 
agree with the band  structure calculations.\cite{Wei,Larson} For such  reasons, the TM 
3$d$ partial spectral weights have been tried to be interpreted in terms of a many-body 
approach like a cluster model calculation with configuration interaction\cite{Mizokawa,CI} 
(CI)  or   an Anderson   impurity  model   calculation.\cite{Gunnarsson} These   model 
calculations, which are  characterized by  phenomenological physical  quantities such  as 
TM 3$d$ on-site Coulomb energy, $p-d$ hybridization, and the ligand $p$ to TM 3$d$ 
charge transfer energy, have been very successful  in describing the electronic structure 
of TM compounds.\cite{TMC}

Soft x-ray absorption  spectroscopy (XAS)  is another  powerful high-energy  probe to 
investigate the electronic structure  of transition metal  compounds.\cite{XAS} Especially 
in diluted  systems, XAS  has big  advantages over  the photoemission   measurements. 
Figure 1 shows XAS spectra at TM $L_{2,3}$-edges of \zms\ ($x  = 0.059$), \zfs\ ($x  
d= 0.049$), and \zcs\  ($x = 0.050$), which  result from TM  2$p$ $\rightarrow$ 3$d$ 
dipole transitions. The  spectra are  dominated by the  large 2$p$  core hole spin-orbit 
coupling energy, which divides them into roughly  the $L_3$ and $L_2$ regions at low 
and high photon energy, respectively. The absorption  process is a local process and its 
energy is determined by  the characteristic core-hole  energy. Despite the fact  that the 
systems contain relatively small concentration of TM atoms, all the spectra nicely reveal 
the complex  spectral  structure, which   is originated  from the  2$p$  core-hole-3$d$ 
Coulomb multiplets, and one  can determine important physical  quantities to understand 
the electronic structure by analysing them. 

The photoemission spectroscopy is known to directly  probe the electronic structure and 
the TM 3$p$-edge resonant photoemission  spectroscopy enables us to extract  the TM 
3$d$ partial spectrum. However, the extracted spectrum  somehow differs from the true 
TM 3$d$ electronic structure because of the  different transition matrix elements in the 
resonant process from those in the direct photoemission process. Hence it is  worthwhile 
to  verify  the  physical  quantities  determined  from   the TM   3$p$-edge resonant 
photoemission studies by  analysing the  XAS results.  Furthermore, in  a view  of the 
experimental technique, the  XAS is  rather bulk-sensitive, different  from the  resonant 
photoemission spectroscopy,  and the  results are   much less sensitive  to the   surface 
condition of  samples. Only  a few   XAS studies, however,  have been  performed  for 
DMS's so far.\cite{Kisiel,Pong1,Oleszkiewicz,Pong2}  Further, the  reported studies were 
mostly focused on XAS  spectra of the ligand  atoms such as  S $K$-edge\cite{Pong1} 
and Te     $L_1$-edge and   $L_3$-edge.\cite{Oleszkiewicz} The  studies  concluded 
qualitatively that the  $p$-$d$ hybridization  is strong  and the $p$-$d$  hybridization 
strength depends on the ligand ions.\cite{Pong2}  However, quantitative analysis for  the 
electronic structure has not been carried out.

In this  paper,  we report  high-resolution  XAS studies   to investigate the   electronic 
structure of various  II-VI DMS's. The  XAS spectra at  the TM $L_{2,3}$  edges are 
analysed by using a tetrahedral cluster model calculation including CI as well as the full 
ionic multiplets. Previously, the model  calculations for TM $L{2,3}$-edge  XAS spectra 
were performed only for condensed systems such as divalent Ni  compounds,\cite{Laan1} 
Ti$_2$O$_3$,\cite{Uozumi} CoO,\cite{Okada} and LiVO$_2$.\cite{Pen},  and showed that 
the CI considerably affects  the XAS spectrum.  Thus it is  also important to  elucidate 
how the effects of CI appear in the XAS spectra of diluted systems like DMS's. 

\section{Experiment}

The samples used  in this study  were \zms\ ($x  = 0.059$), \zfs\  ($x = 0.049$),  and 
\zcs\ ($x =  0.050$) thin films  grown by molecular  beam epitaxy on  the GaAs (001) 
substrate and a \zmt\ ($x = 0.60$) single crystal with (110)  surface. The details of the 
sample  growth  and  characterization  were  reported  elsewhere.\cite{Jonker}  All  the 
samples preserve the zinc-blende crystal structure. The XAS measurements were carried 
out at  the  Dragon high-resolution  soft-x-ray  beam line\cite{Chen}  at  the National 
Synchrotron Light Source,  Brookhaven National  Laboratory. The  energy resolution  of 
incoming photon  was set   to be about  0.5~eV  in the  full width   at half maximum 
(FWHM). Before the measurements, the  single-crystalline sample was cleaved  {\em in 
situ} and  the thin-film   samples, which had   been pre-etched using   1:3 mixture of 
NH$_4$OH (29~\%):methanol,\cite{Etching} were annealed {\em in situ} at $300~^\circ$C 
by radiation  heating for  about  three hours.\cite{Annealing}  The measurements  were 
performed  at  room   temperature and   the  base  pressure   was maintained   below 
2$\times10^{-10}$~Torr.

\section{Calculational Method}

A tetrahedral cluster model  calculation, which we  have employed to analyse  the XAS 
spectra, includes not only the full multiplets of TM 3$d$ electrons but also configuration 
interaction. The model is characterized  by the on-site 3$d$ Coulomb  energy, $U\equiv 
E(d^{n+1})+E(d^{n-1})-2E(d^{n})$,    the    ligand-to-3$d$   charge-transfer    energy, 
$\Delta\equiv    E(d^{n+1}\underline{L})-E(d^n)$,   and    the    cation-$d$-ligand-$p$ 
hybridization strength $V$. The  initial configuration states are  spanned over the  ionic 
ground  state   $3d^n$ and   the  charge-transferred   states, $3d^{n+1}\underline{L}$, 
$\cdots$, $3d^{10}\underline{L}^{10-n}$,  where  $\underline{L}$ denotes  a  ligand-$p$ 
hole. The multiplets of the $3d$ electron configurations in  the charge transferred states 
are taken into account explicitly while the  total symmetry is preserved to be the  same 
as the  ionic ground  state  symmetry. The  3$d$ Coulomb  exchange  interactions are 
represented by Racah  parameters, $B$ and  $C$. Although the  Racah parameters can 
vary   slightly   in   different   configurations,    we   imported   the   values   from 
Ref.~\onlinecite{Mizokawa}  and  fixed  them  in  all  configurations  for  the  sake  of 
simplicity. The  on-site Coulomb  $U$ and   the charge transfer  energy $\Delta$   are 
treated as control parameters in the calculation and defined with the lowest-energy $3d$ 
multiplet states of the corresponding configurations. 

Similarly,   the   final   states  are   also   spanned   over   $\underline{2p}3d^{n+1}$, 
$\underline{2p}3d^{n+2}\underline{L}$,                                         $\cdots$, 
$\underline{2p}3d^{10}\underline{L}^{9-n}$ configuration  states. The   Coulomb energies 
between the 2$p$  core-hole and  a 3$d$  electron are represented  by the  2$p$-3$d$ 
Slater integrals, $F^2_{pd}$, $G^1_{pd}$, and $G^3_{pd}$, and the spin-orbit coupling of 
the $2p$ core-hole is  taken into account  with ($\zeta_{2p}$). The involved  parameter 
values are borrowed from Ref.~\onlinecite{Laan2} and  fixed for all  configurations. The 
Slater integrals were scaled  down by a reduction  factor $\kappa$ in order  to account 
for the   solid state   screenings.\cite{Lynch} The   average Coulomb  interaction  $Q$ 
between a  TM 2$p$  hole and  a 3$d$  electron is  set to  be an  empirical value  of 
$1.25U$ as widely adopted in  the core-level spectroscopy studies. 

The hybridization interactions are  expressed in terms  of the Slater-Koster  parameters 
$V_{pd\sigma}$ and   $V_{pd\pi}$: $V_{t}\equiv   \langle t   | H  |  L_{t}  \rangle  = 
\sqrt{\frac{4}{3}V_{pd\sigma}^2 + \frac{8}{9} V_{pd\pi}^2}$ and $V_e \equiv \langle  | H 
| L_e \rangle = \frac{2\sqrt{6}}{3} V_{pd\pi}$, where $t$($e$)  and $L_{t}$ ($L_e$) are 
TM-$3d$  and  ligand-$p$ orbitals  with $t_{2g}$  ($e_g$) symmetry  orbitals under 
$T_d$ tetrahedral  point group  symmetry, respectively.  For  simplicity, we  apply the 
empirical  relation  $V_{pd\sigma}  =  -2  V_{pd\pi}$.\cite{Harrison}  The  crystal-field 
interaction 10$Dq$ is fixed to be 0.25~eV in all the  calculations, and the reason will be 
explained later. Here TM  3$d$ spin-orbit coupling  is neglected. It  is known that  the 
3$d$ spin-orbit coupling can  be quenched when the  ground-state orbital symmetry  is 
either $A$ or $E$\cite{Laan2}, and indeed, the  ground state symmetries of Mn$^{2+}$, 
Fe$^{2+}$, and Co$^{2+}$  ions are $^6A_1$,  $^5E$, and $^4A_2$  under the   $T_d$ 
symmetry, respectively.

The initial ground  state is obtained  in the modified  Lanczos method.  Then the XAS 
spectra are calculated by the  continued fraction expansion.\cite{Dagotto} The  calculated 
spectrum is broadened  with a  Lorentzian function,  which accounts  for the  core-hole 
lifetime, and finally with  a Gaussian function  for the experimental  resolution (0.5~eV). 
The  Lorentzian  broadenings,  $\Gamma_2$  for  $L_2$-edge  and   $\Gamma_3$ for 
$L_3$-edge spectra are presented in Table I.

\section{Results and Discussion}

\subsection{CI effects on the cluster model calculation}

The calculated   TM $L_{2,3}$-edge   XAS spectra   for Mn$^{2+}$,   Fe$^{2+}$, and 
Co$^{2+}$ ions,   which are  in a   tetrahedral cluster,  are presented   in Figure~2  in 
comparison with the corresponding  experimental XAS spectra  of \zms\ ($x  = 0.059$), 
\zfs\ ($x = 0.049$),  and \zcs\ ($x  = 0.050$), respectively.  Now the photon  energy is 
presented relative to  that of  the corresponding  main $L_3$  multiplet peak  with the 
highest intensity.   In order  to investigate   the role  of configuration   interaction, the 
calculation has been performed for the orders of the charge transferred states. 

In the zero-th  order calculations, which  are denoted by  $\underline{L^0}$, the charge 
transferred states are neglected, $i.e.$  the ionic calculations. Meanwhile in  the $m$-th 
order calculations,  denoted  by $\underline{L^m}$,   the charge  transferred states  are 
included up to  the configuration states  with $m$ ligand  holes, $i.e.$  the calculations 
with   the  configuration   states   of   $3d^n$,  $3d^{n+1}\underline{L}^1$,   $\cdots$, 
$3d^{n+m}\underline{L}^m$  for  the  initial  states  and  the  configuration   states of 
$\underline{2p}3d^{n+1}$,        $\underline{2p}3d^{n+2}\underline{L}^1$,       $\cdots$, 
$\underline{2p}3d^{n+m+1}\underline{L}^m$  for   the final   states.   The spectra   are 
normalized by the intensity  of the $L_3$  main peak. For  a given TM  ion, the same 
parameter set is  employed for  the $\underline{L^0}$, $\cdots$,  the $\underline{L^m}$ 
calculations.

When we take  into account  only the $\underline{L}^1$  charge transferred  states, we 
find that   the calculated  spectrum changes   considerably from  the $\underline{L^0}$ 
spectra, especially in the cases of Fe$^{2+}$ and Co$^{2+}$ ions.\cite{comment} But the 
$\underline{L^2}$  spectra  are  nearly  the  same  as  the  $\underline{L}^1$  spectra, 
indicating that the CI effects on  the spectrum converge very rapidly.  It is because the 
energies of  the  $3d^{n+2}\underline{L}^2$ and  $\underline{2p}3d^{n+3}\underline{L}^2$ 
configuration    states    already    become     much    higher    than    those    of 
$3d^{n+1}\underline{L}^1$ and  the  $\underline{2p}3d^{n+2}\underline{L}^1$ states,   due  
to the strong on-site Coulomb interaction. Thus we will consider, from now on, only the 
result of calculations  which include configurations  up to  $d^{n+2}\underline{L}^2$ and 
$\underline{p}d^{n+3}\underline{L}^2$. Here the parameter values are chosen to give  the 
$\underline{L^2}$ spectra, which agree well with the corresponding experimental ones.

As can be seen in Fig.~2,  the spectrum is strongly disturbed by  CI, especially in high 
energy region of the main $L_3$  multiplet peak. The peak separation becomes  reduced 
and the sharp peak structure is  smeared out with CI. Although similar  trend might be 
reproduced just by increasing the crystal  field splitting 10$Dq$ value without including 
CI,\cite{Laan2} we believe CI  is essential for proper  interpretation of the XAS  spectra 
for the following reason. When we  increase the value of 10$Dq$  without including CI, 
the pre-edge structure is well separated from the main  peak and grows rapidly. At the 
same time the $L_2$-edge spectrum becomes very  complicated except for a Co$^{2+}$ 
ion,\cite{deGroot} neither  of  which is  observed  in the  experimental  spectra. Hence, 
10$Dq$ value  should be  small for  Mn$^{2+}$ and   Fe$^{2+}$ ions but  large for  a 
Co$^{2+}$ ion if one  fits the experimental spectra  without CI. It is,  however, hard to 
believe that   the 10$Dq$   value, which   reflects the   hybridization strength  in  the 
calculation, changes abruptly along this series.\cite{Park} 

In the comparison of the calculated spectra with the experimental ones, the agreement is 
rather good in overall. But  one can still recognize  some minor disagreement, which  is 
expected to  arise from  the uncertainties   in Slater integrals  and spin-orbit   coupling 
parameters and the solid  state effects which  have been left  out in the cluster  model. 
Each final state multiplet  may have different lifetime,  and thus for  the better fit, one 
should apply an appropriate core-hole-lifetime  broadening for each multiplet  state. The 
solid state effects seem  to be rather severe  in the spectrum  of a Co$^{2+}$ ion,  and 
thus we employed the reduction factor $\kappa$ smaller than other TM ions. 

The parameter values  for the  best fit  are tabulated in  Table~1. The  ligand-to-3$d$ 
charge transfer energy  $\Delta$ of Mn,  Fe, and Co  DMS's are  5.0, 3.0, and  2.5~eV, 
respectively. Here $\Delta$ is defined  with respect to the  center of the multiplet.  The 
decreasing behavior of $\Delta$ value along the  series agrees with the decrease of  the 
difference in the  electro-negativity between  TM and ligand  atoms. These  values are 
similar to them obtained from the $d$-$d$ optical-absorption study.\cite{Mizokawa} The 
hybridization parameter $V_{pd\sigma}$  are determined to  be 1.0, 0.9,  and 0.8~eV for 
Mn, Fe, and Co DMS's,  respectively. The small decrease  of $V_{pd\sigma}$ with the 
increasing atomic number is also  consistent with the expected  contraction of the 3$d$ 
electron wave function, but these values are somewhat smaller than those obtained from 
the analysis  of photoemssion  spectra.\cite{Mizokawa,CI} This  is probably  due to  the 
small reduction of the  effective hybridization strength in  the XAS final states  induced 
by the strong Coulomb potential by the localized core orbitals as pointed by Gunnarsson 
and Jepsen.\cite{Jepsen} Similar  behavior is  also observed  in the  analysis of  the Ni 
compounds using the Anderson impurity model.\cite{Laan1}

\subsection{Hybridization effect}

In order to study the details of  the hybridization effect, we compared the  XAS spectra 
of \zms\ and \zmt\  where different ligands will  give variations in the  charge-transfer 
energy $\Delta$ and the  hybridization interaction $V_{pd\sigma}$.  As can be  seen in 
Fig.~1, the Mn $L_3$-edge XAS spectrum shows a simple and distinguishable multiplet 
structure, and thus is very appropriate to study the variation of the hybridization  effect. 
Figure~3 shows   the experimental  XAS spectra   of Zn$_{0.941}$Mn$_{0.059}$Se  and 
Zn$_{0.4}$Mn$_{0.6}$Te  at  the  Mn  $L_3$-edge  together  with  the  corresponding 
calculated ones.\cite{zms} The experimental spectra show four  peak structure, a highest 
intensity peak (denoted  by $A$), two  high energy peaks  at about 1~eV  (denoted by 
$B$) and 3.5~eV (denoted by $C$)  above, and a small pre-edge  structure (denoted by 
$D$),  for  both  compounds.   This structure   is  mainly  originated  from  the   Mn 
2$p^5$3$d^6$ XAS final state multiplets of the Mn$^{2+}$ ion. 

The spectra   display the  common  multiplet  structure, but   one can  still  recognize 
considerable difference   in the   relative peak   energies and   intensities for  different 
ligand-ions. For examples,  the relative energy  of the  peak B is  lower for Te-ligand 
ions, while the width of  the peak C is narrower  for the Se-ligand ions. The  best fits 
for the multiplet of these  two compounds were obtained  with the parameter values  of 
$\Delta$ = 5.0~eV and $V_{pd\sigma}$  = 1.00~eV for Zn$_{0.941}$Mn$_{0.059}$Se and 
$\Delta$ = 3.8~eV  and $V_{pd\sigma}$ =  0.90~eV for Zn$_{0.4}$Mn$_{0.6}$Te.  Other 
parameters such as the on-site Coulomb interaction $U$ and the crystal-field interaction 
10$Dq$ are fixed to be 8.0~eV and 0.25~eV,  respectively, since they are expected to be 
hardly affected  by  the variation   of the  ligand ions.   In the  picture, $\Delta$  and 
$V_{pd\sigma}$ mainly  determine  the hybridization   strength between  TM-$d$ and 
ligand-$p$ orbitals. 

As can be seen in the figure, the calculated spectra show good over all agreement with 
the experimental ones. Chemical trends of the  parameter values are consistent with the 
electro-negativity of ligand ions  and also with the  nearest neighbor distances between 
Mn and ligand ions. According to the Harrison's scheme,\cite{Harrison} the bond-length 
($d$) dependence   of $V_{pd\sigma}$  is $d^{-3.5}$,   but the  obtained values   yield 
$d^{-1.4}$ dependence using a  simple central-force model\cite{Shih}  for the Mn-anion 
bond length. Such  a disagreement, which  was also observed  by Larson  {\em et al.}, 
may be attributed to the chemical difference  between Te and Se ions.\cite{Larson} The 
agreement in the  region of  the second satellite  is not  as good as  the first  satellite, 
especially in  Zn$_{0.4}$Mn$_{0.6}$Te. We  suspect that  this is  probably  due to  the 
magnetic interaction between  Mn$^{2+}$ ions  in this  Mn-rich compound,  which can 
affect the spectral shape considerably.\cite{Veenendaal}

It is worthwhile to  understand how the  XAS spectrum is  affected by the parameters 
$\Delta$ and $V_{pd\sigma}$. Fig.~4 shows the XAS spectra as a  function of $\Delta$ 
for a fixed $V_{pd\sigma}$ = 1.0~eV in the left  panel and the spectra as a function of 
$V_{pd\sigma}$ for a fixed $\Delta$ =  4.0~eV in the right panel.  First, the calculation 
results show that  the position  of the  peak $B$  is closer to  the peak  $A$ and  its 
intensity becomes   smaller when  the  hybridization effect   increases by  decrease  of 
$\Delta$ or increase of $V_{pd\sigma}$. Second, the  structure $C$ in the experimental 
spectrum seems to  be a  single peak  structure. However,  it turns  out to  consist of 
several multiplet states.  Third, the pre-edge  structure $D$ is  mainly induced  by the 
crystal-field splitting 10$Dq$.  This structure  is absent  at 10$Dq  = 0$  but becomes 
distinguishable for  large 10$Dq$-value.\cite{Laan2}  In  our parameter  set with  small 
10$Dq$, the  structure $D$  is  barely recognized  as in  the  experimental results.  In 
practice, it is difficult to determine the  exact value of 10$Dq$ from the fitting  of XAS 
spectra, and we simply fixed 10$Dq$ = 0.25~eV just enough to show the structure $D$.

\section{Conclusion}

XAS spectra of  various DMS's at  the TM  atom $L_{2,3}$ edges  are presented and 
described  by  a  tetrahedral   cluster model   including full   multiplet  structure  and 
configuration interaction. Due  to the  strong hybridization  between the  TM-3$d$ and 
ligand-$p$ orbitals, inclusion  of CI  with reasonable  physical parameters  gives much 
better results than  the model without  CI. Thus  it is important  to incorporate CI  for 
understanding of the XAS spectra of DMS's.  Hybridization effects on XAS spectra are 
also investigated varying  ligand ions  of Mn DMS's,  and the  change of  the spectral 
shape is well explained by the cluster model with CI.

\acknowledgments

{This work was supported by the Korean Science and Engineering Foundation (KOSEF) 
through the  Center  for Strongly   Correlated Materials Research   (CSCMR) at  Seoul 
National  University   (2000), and   Grant  No.   KOSEF 951-0209-014-1.   J.-S.  Kim 
acknowledges the support from Atomic Scale Surface Science Research Center (ASSRC) 
at Yonsei  University. The  work is  also supported  by BK21  project of   Ministry of 
Education, Korea (SNU,  POSTECH, and  Kyung Hee  U.). National Synchrotron  Light 
Source, Brookhaven  National Laboratory,  at which  the  XAS data  were obtained,  is 
supported by  the US  Department of  Energy. We  also thank  Prof. M.  Taniguchi of 
Hiroshima University, Japan, for providing a Zn$_{0.4}$Mn$_{0.6}$Te single crystal.}

\begin{table}
\caption{Parameters used  to  calculate TM  $L_{2,3}  $XAS spectra  of  \zms\ ($x  = 
0.059$), \zfs\ ($x 
= 0.049$), and \zcs\ ($x = 0.050$). See text for detailed explanations. All the values are 
given in eV.} 
\vspace{8mm}
\begin{tabular}
{cccccccccccccc}  &$\Delta$&$V_{pd\sigma}$&  $U$   & $Q$   &  $B$  &   $C$ & 
$F^2_{pd}$ &  $G^1_{pd}$&$G^3_{pd}$ &  $\zeta_{2p}$  & $\kappa$&$\Gamma_2$  & 
$\Gamma_3$ \\ \tableline  Mn$^{2+}$ & 5.0  & 1.00 &  8.0 &10.0 &  0.119 &  0.412 & 
6.321 & 4.606 & 2.618 &   6.846 & 0.8  & 0.45 & 0.35 \\Fe$^{2+}$ & 3.0 &  0.90 & 3.6 
& 4.5 &  0.131 &  0.484 & 6.793  &   5.004  & 2.844 &  8.200 &  0.8 &  0.70 & 0.45 
\\Co$^{2+}$ & 2.5 &  0.80 & 4.8 &  6.0 & 0.138 &  0.541 & 7.259  & 5.397 & 3.069  & 
9.748 & 0.7 & 1.00 & 0.50\end{tabular}
\end{table}

\begin{figure}
\caption{Experimental XAS spectra  of \zms\  ($x = 0.059$),  \zfs\ ($x  = 0.049$),  and 
\zcs\ ($x = 0.050$) at the TM $L_{2,3}$ edges.}
\end{figure}

\begin{figure}
\caption{Experimental (dots) and Calculated (lines) $L_{2,3}$ XAS spectra of Mn$^{2+}$,  
Fe$^{2+}$, and Co$^{2+}$  ions in  a tetrahedral  cluster depending  on the  number of 
configurations employed.  $\underline{L}^n$ denotes  a configuration,  which contains  a 
$\underline{L}^n$ term in the initial  and final states, up to  which was included in the 
calculations. See text for detailed information on employed parameter values. }
\end{figure}

\begin{figure}
\caption{ Experimental (dots) and calculated (lines) XAS spectra of \zms\ ($x = 0.059$) 
and  \zmt\ ($x =  0.60$) at the Mn  $L_3$ edge. The vertical  line is a guide  for the 
peak position. For parameter values of the calculations, refer to text. }
\end{figure}

\begin{figure}
\caption{ Calculated $L_3$  XAS spectra of  a Mn$^{2+}$ ion  in a tetrahedral  cluster. 
Left panel: As a function of  $\Delta$. $V_{pd\sigma}$ is fixed to  1.0~eV. Right panel: 
As a function of  $V_{pd\sigma}$. $\Delta$ is  fixed to 4.0~eV.  The vertical line  is a 
guide for the peak position.}
\end{figure}


\begin{references}
\bibitem[*] {address1}Author to whom all correspondence should be addressed.
\bibitem{zms} We could  not find any  noticeable difference  in the Mn  $L_{2,3}$-edge 
XAS spectrum for different x in \zms\ (x=0.059, 0.1).
\bibitem{comment} For a  Mn$^{2+}$ ion, the  effect is  still noticeable but  small. The 
reason might be  that the  Mn$^{2+}$ ion  has relatively simple  multiplet structure  in 
comparison with other ions.
\end{references}
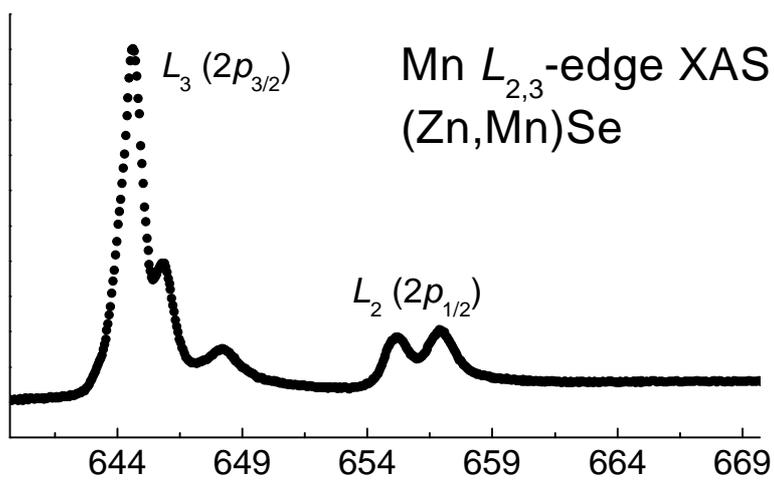
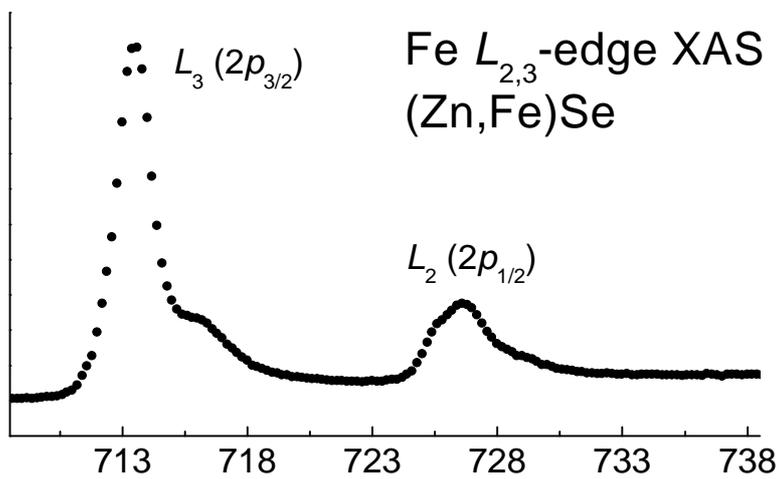
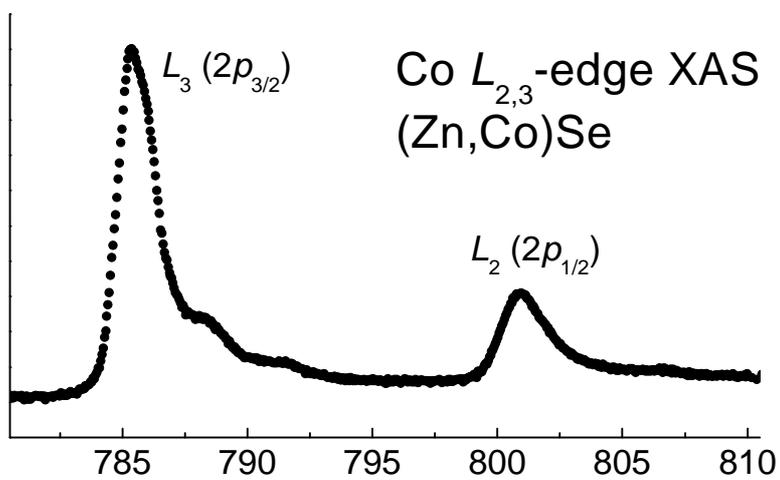
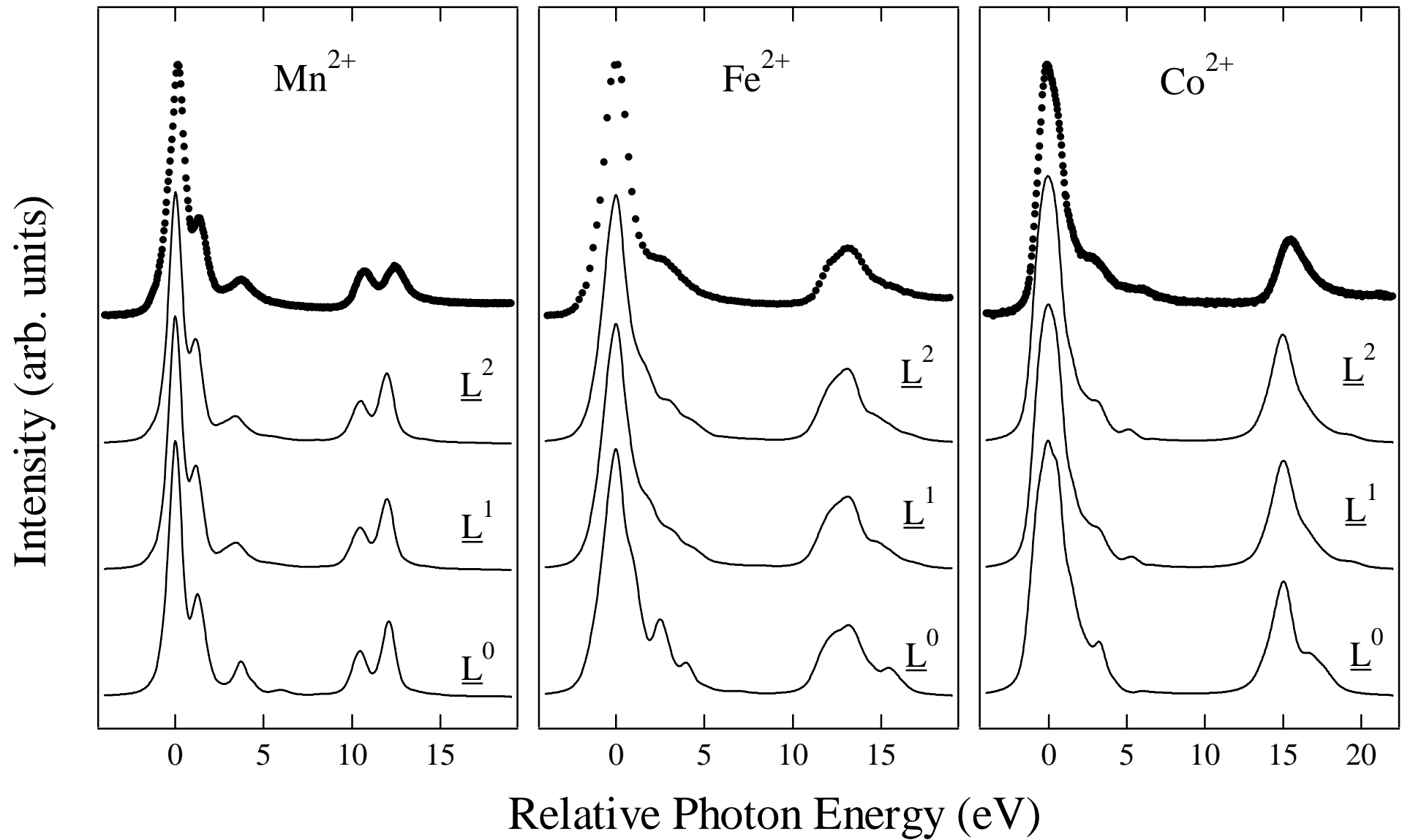
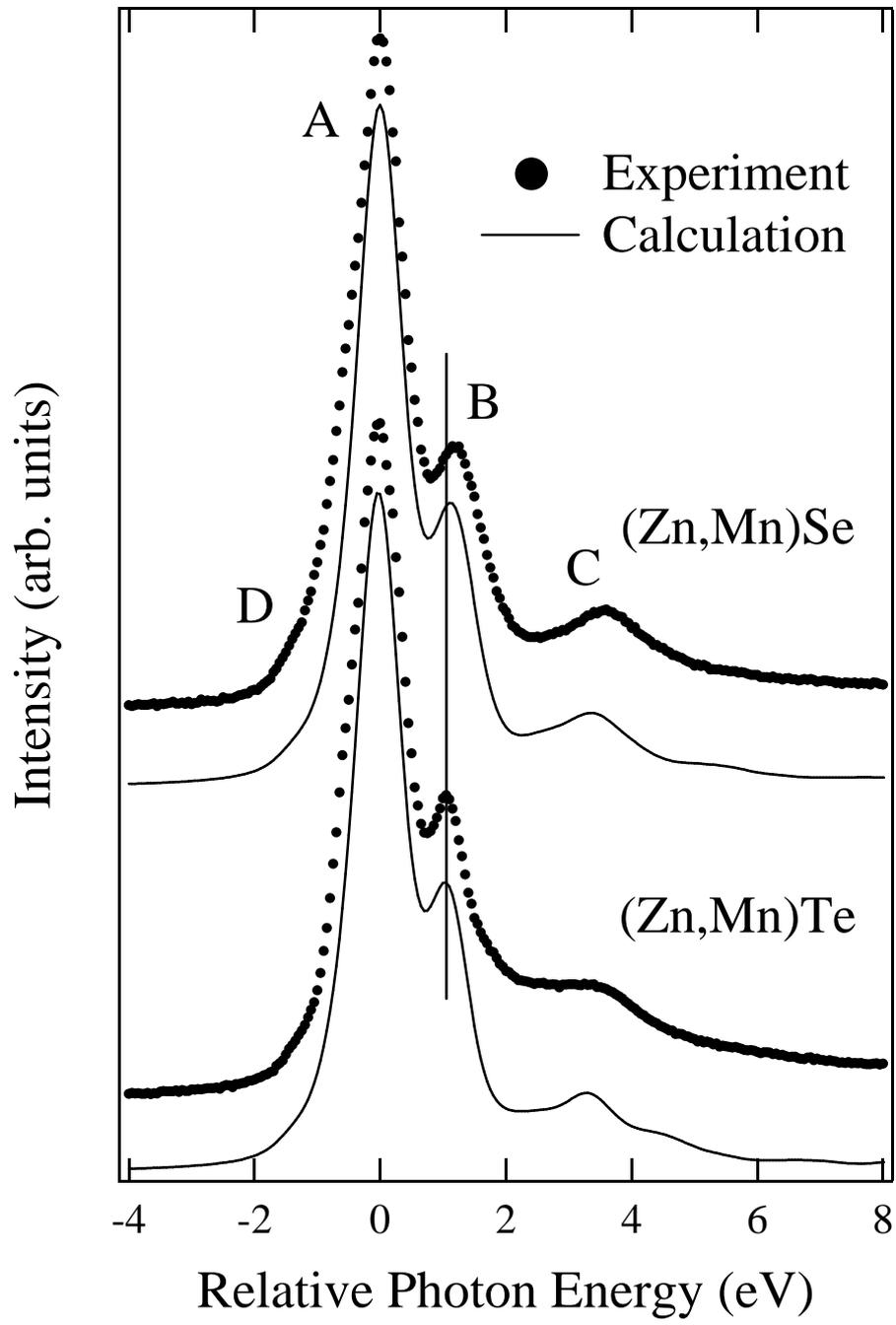
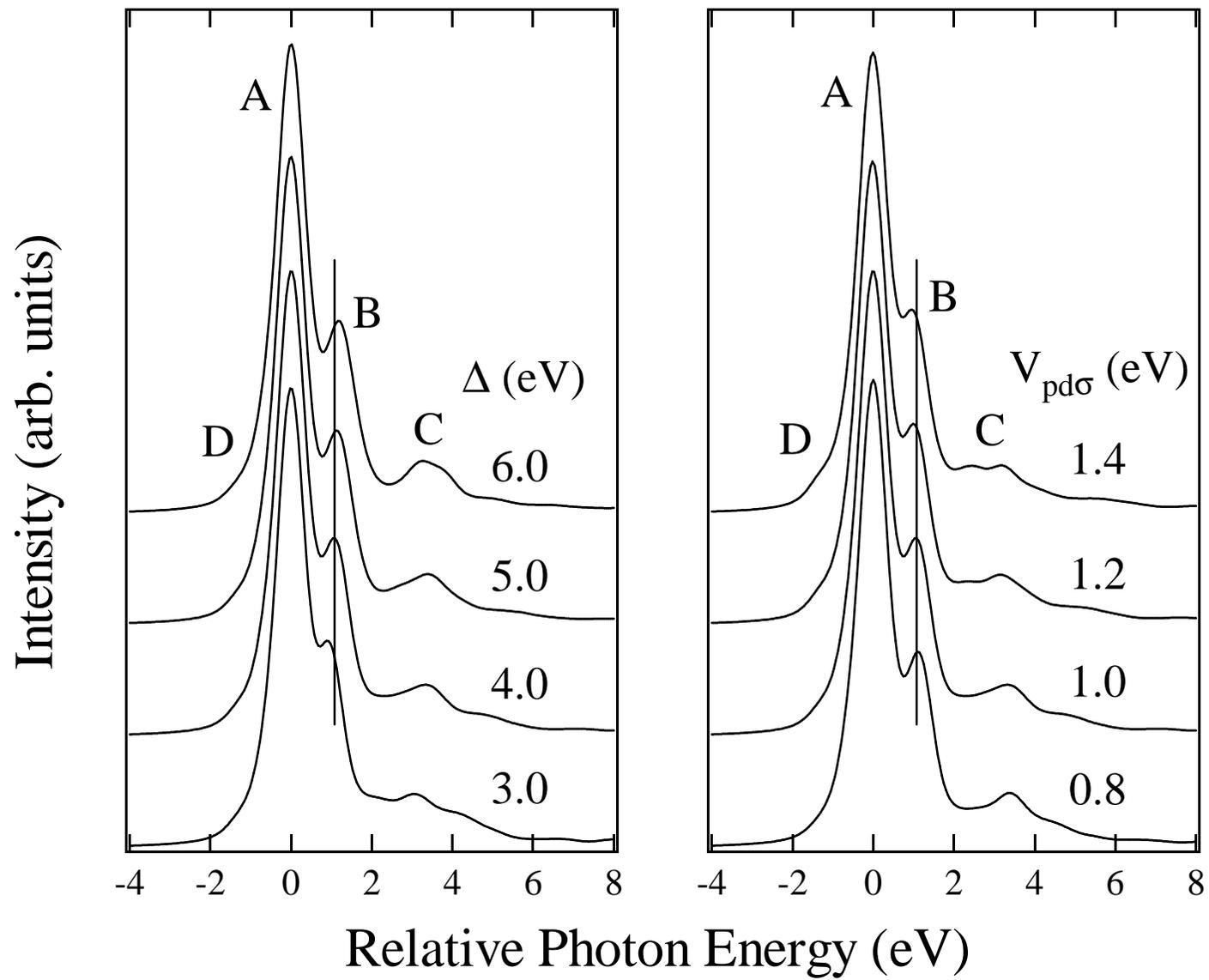
\end{document}